\documentclass[fleqn,10pt]{wlscirep}
\makeatletter
  \if@titlepage
  \renewcommand\maketitle{\begin{titlepage}%
  \let\footnotesize\small
  \let\footnoterule\relax
  \let \footnote \thanks
  \null\vfil
  \vskip 60\p@
  \begin{center}%
    {\LARGE \@title \par}%
    \vskip 3em%
    {\large
     \lineskip .75em%
      \begin{tabular}[t]{c}%
        \@author
      \end{tabular}\par}%
      \vskip 1.5em%
    {\large \@date \par}%       % Set date in \large size.
  \end{center}\par
  \@thanks
  \vfil\null
  \end{titlepage}%
  \setcounter{footnote}{0}%
  \global\let\@thanks\@empty
  \global\let\@author\@empty
  \global\let\@date\@empty
  \global\let\@title\@empty
}
\else
\renewcommand\maketitle{\par
  \begingroup
    \renewcommand\thefootnote{\@fnsymbol\c@footnote}%
    \def\@makefnmark{\rlap{\@textsuperscript{\normalfont\@thefnmark}}}%
    \long\def\@makefntext##1{\parindent 1em\noindent
            \hb@xt@1.8em{%
                \hss\@textsuperscript{\normalfont\@thefnmark}}##1}%
    \if@twocolumn
      \ifnum \col@number=\@ne
        \@maketitle
      \else
        \twocolumn[\@maketitle]%
      \fi
    \else
      \newpage
      \global\@topnum\z@   % Prevents figures from going at top of page.
      \@maketitle
    \fi
    \thispagestyle{plain}\@thanks
  \endgroup
  \setcounter{footnote}{0}%
  \global\let\@thanks\@empty
  \global\let\@author\@empty
  \global\let\@date\@empty
  \global\let\@title\@empty
}
\makeatother
\usepackage{caption,subcaption}
\title{Abrupt current switching in graphene bilayer tunnel transistors enabled by van Hove singularities}

\author[1,2]{Georgy Alymov}
\author[1,2]{Vladimir Vyurkov}
\author[3]{Victor Ryzhii}
\author[1,2,*]{Dmitry Svintsov}

\affil[1]{Department of Physical and Quantum Electronics, Moscow Institute of Physics and Technology, Dolgoprudny 141700, Russia}
\affil[2]{Laboratory of Sub-micron Devices, Institute of Physics and Technology  RAS, Moscow 117218, Russia}
\affil[3]{Research Institute of Electrical Communication, Tohoku University, Sendai 980-8577, Japan}

\affil[*]{svintcov.da@mipt.ru}

\begin{abstract}
In a continuous search for the energy-efficient electronic switches, a great attention is focused on tunnel field-effect transistors (TFETs) demonstrating an abrupt dependence of the source-drain current on the gate voltage. Among all TFETs, those based on one-dimensional (1D) semiconductors exhibit the steepest current switching due to the singular density of states near the band edges, though the current in 1D structures is pretty low. In this paper, we propose a TFET based on 2D graphene bilayer which demonstrates a record steep subthreshold slope enabled by van Hove singularities in the density of states near the edges of conduction and valence bands. Our simulations show the accessibility of $3.5\times10^4$ ON/OFF current ratio with 150 mV gate voltage swing, and a maximum subthreshold slope of (20 $\mu$V/dec)$^{-1}$ just above the threshold. The high ON-state current of  $0.8$~mA/$\mu$m is enabled by a narrow ($\sim 0.3$ eV) extrinsic band gap, while the smallness of the leakage current is due to an all-electrical doping of the source and drain contacts which suppresses the band-tailing and trap-assisted tunneling.
\end{abstract}
\begin{document}

\flushbottom
\maketitle
\thispagestyle{empty}
\section*{Introduction}
The design of field-effect transistors (FETs) operating at sub-$0.5$ V supply voltage is one of the major challenges for nanoelectronics paving the way to resolve the problem of power dissipation in large integrated circuits. Tunnel FETs (TFETs) with interband tunneling are among the principal candidates to meet this demand~\cite{Ionescu_tunnel,Seabaugh_State_of_Art}. The low-voltage switching in TFETs occurs due to a sharp dependence  of the tunnel current on the conduction-valence band overlap in a gate-controlled $p$-$i$ or $p$-$n$ junction~\cite{Seabaugh_Low-voltage}; once there is no band overlap, there is no tunnel current. This fact, alongside with the smallness of thermionic leakage current and efficient modulation of the barrier transparency by the gate voltage, results in the subthreshold slope of the TFET characteristics surpassing the thermionic limit~\cite{Choi,Tomioka_InAsNWTFET,krishnamohan,Appenzeller,HannaSREPInAs} of ($60$ mV/dec)$^{-1}$.

It is intuitive that the abrupt variations of electron and hole densities of states (DoS) near the band edges further enhance the switching efficiency of TFETs~\cite{agarwal_DOS_switch}. For a $d$-dimensional TFET channel, the DoS scales with energy $E$ above the band edge as $E^{(d-2)/2}$, while the dependence of current $J$ on the gate voltage $V_G$ above the threshold $V_T$ is $J\propto [V_G - V_T]^{(d+1)/2}$~\cite{Kane_Theory_of_tunneling, Keldysh}. Apart from the density-of-states-enhanced switching, the TFETs with low-dimensional channels demonstrate an improved electrostatic control of the band alignment by the gate voltage~\cite{Knoch_SSE,HannaSREPInAs,Fahad_SiNT}. Theoretically, the effects of DoS on current switching steepness are most pronounced in vertical TFETs based on the two-dimensional crystals~\cite{britnell_resonant,2GBL_vertical_exp} and electron-hole bilayers in quantum-confined structures~\cite{EH-bilayer,agarwal_bilayer_tunnel}. In such TFETs, the joint density of states is nonzero just at one certain value of gate voltage~\cite{SymFet} -- which could lead to the abrupt-most current switching ever. In practice, the density-of-states effects on the subthreshold steepness are largely smeared. The reason for the smearing in vertical TFETs based on van der Waals heterostructures is the rotational misalignment of 2D layers in momentum space~\cite{twist-controlled}. In common semiconductor structures it is believed that charged defects and dopants lead to pronounced band-tailing and emerging trap-assisted and band-tail tunneling leakage currents~\cite{Vandooren_SSE,Trap-Assisted,Khayer_JAP}.

\begin{figure}[ht]
\includegraphics[width=1.0\linewidth]{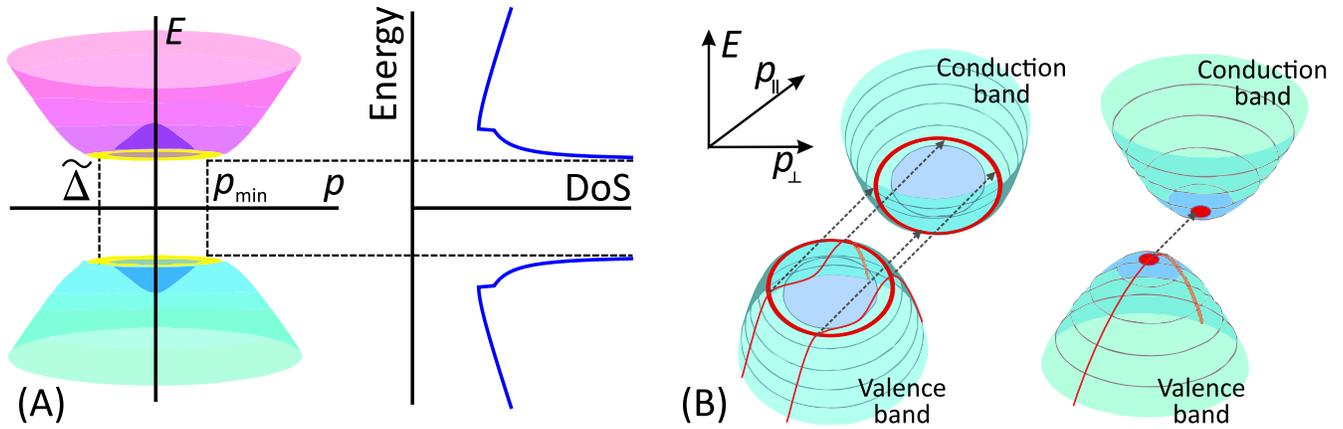}
\caption{\label{F1} (A) Electron spectrum $E(p)$ in graphene bilayer under transverse electric field and the energy dependence of its DoS. The ''Mexican hat'' feature in the dispersion law leads to the square-root singularities in the DoS near the band edges. Panel (B) highlights with red the electron states involved in the interband tunneling at small band overlap in graphene bilayer (left) and in a semiconductor with parabolic bands (right). The phase space for tunneling in graphene bilayer represents a ring, while in a parabolic-band semiconductor it is a point. Dashed lines indicate the tunneling transitions, red lines indicate the trajectories of the tunneling electrons in the valence band. 
}
\end{figure}

In this paper, we theoretically demonstrate that graphene bilayer (GBL) represents an ideal platform for the low-voltage tunnel switches. A peculiar 'mexican hat' band structure of GBL formed under transverse electric field~\cite{McCann_EPJ,Zhang_gap} leads to a van Hove singularity in the DoS right at the bottom of the band edges, as shown in Fig.~\ref{F1} A. It was experimentally proved that this singularity leads to an enhanced infrared absorption~\cite{Kuzmenko_PRB}. Another manifestations of this singularity predicted theoretically include a step-like change of the GBL FET channel conductance~\cite{Svintsov_JJAP} and multi-peak structure of the vertical GBL TFET characteristics~\cite{2GBL_vertical_theory}. In this paper we show that such singularity results in a steep, linear dependence of the GBL TFET current on the gate voltage above the threshold, which was attributed previously just to the TFETs based on one-dimensional materials~\cite{Tomioka_InAsNWTFET,Bjork_APL_SiNWTFET,Jena_GNR_the,Jena_1DvanHove,HannaSREPInAs}. 

The advantage of graphene bilayer TFETs over those based on 2d materials with parabolic bands in terms of switching steepness can be illustrated by Fig.~\ref{F2} B. As the conduction and valence bands in a GBL tunnel junction overlap, the electrons capable of tunneling are located on a ring in the momentum space. In contrast, the electrons capable of tunneling between simple parabolic bands are located in a small vicinity of an extreme point of the dispersion. To fully exploit the density-of-states effect for tunneling and get rid of the dopant-induced band edge smearing, we introduce the TFET based on electrical doping by auxiliary gates. Apart from avoiding the parasitic tunnel currents, this adds the possibility  to electrically reconfigure the device between $n$- and $p$-types. Under optimal gate biasing conditions, the proposed TFET demonstrates the current switching over more than $4$ orders of magnitude with 150 mV gate voltage swing only. At the same time, the ON-state current density as large as $0.8$~mA/$\mu$m is accessible due to the low extrinsic band gap of GBL ($\sim 0.3$ eV) and large DoS far above the band edges.

\section*{Results}

\subsection*{Device structure}

\begin{figure}
\includegraphics[width=1\linewidth]{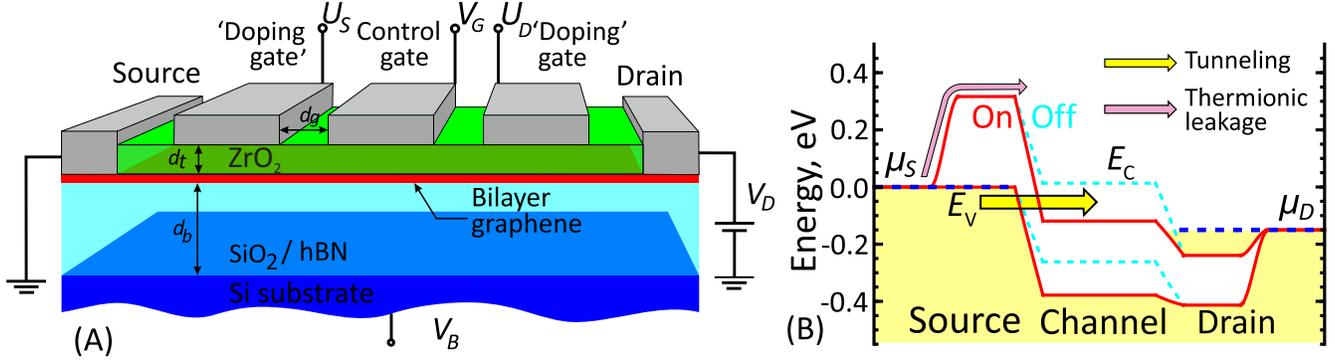}
\caption{\label{F2} (A) Layout of the proposed graphene bilayer TFET with electrically defined source and drain regions (B) Band diagram of graphene bilayer TFET for the optimal biasing conditions: $V_B >0$, $U_S < 0$, $U_D > 0$. At zero top gate bias, $V_G = 0$, the TFET is switched on, while at $V_G <0 $ it is switched off }
\end{figure}

The advantages of graphene bilayer for the steep current switching can be fully realized in the structure of the TFET shown schematically in Fig.~\ref{F2} A. A heavily doped silicon substrate acts as a bottom gate used to create the transverse electric field and thus open and manipulate the band gap in GBL~\cite{Tunable_Gap}. The oxidation of the substrate results in formation of SiO$_2$ layer playing the role of back gate oxide and substrate for graphene bilayer. Alternatively, the SiO$_2$ layer can be replaced with hexagonal boron nitride possessing a small ($\sim10^{10}$ cm$^{-2}$) density of residual charged impurities~\cite{GBL_encapsulated_in_BN}. A nanometre -- thin layer of high-$\kappa$ dielectric (e.g., zirconium oxide) covers the graphene channel, and the top metal gates are formed above. The side gates near the source and drain contacts induce large densities of holes and electrons, respectively, which also leads to the formation of an abrupt tunnel junction and energy barriers~(see Fig.~\ref{F2} B) for the thermally activated electrons and holes contributing to the OFF-state leakage current. 

The operation of a normally open TFET switched off by a negative top gate voltage is illustrated in the band diagram, Fig.~\ref{F2} B. Application of positive voltage to the bottom gate, $V_B > 0$, induces the band gap and provides an excess electron density in bilayer. The $p^+$ doping of source emerges upon application of negative voltage $U_S < 0$ to the source doping gate. An additional increase in the barrier height for the holes injected from the drain is achieved by applying positive voltage $U_D > 0$ to the drain doping gate. It is instructive that application of high voltage to the doping gates does not result in increased power consumption as this voltage is not changed during the device operation. At zero top gate voltage, the valence band in $p^+$--source overlaps with the conduction band in the $n$--type channel, which corresponds to the ON state (red band profiles in Fig.~\ref{F2} B). Upon application of negative voltage to the top gate, the transistor is switched off (dashed blue band profiles in Fig.~\ref{F2} B).

The optimization of the device dimensions aiming at the increase in the ON-state and reduction in the OFF-state currents is quite straightforward: both the effective thickness of the gate dielectric and the distance between the source doping gate and the control gate should be small. These distances are limited just by the possible gate leakage current (see below), we choose them to be $d_t = 2$ nm and $d_g = 5$ nm. The doping gate at the drain is used just to induce high barrier for thermally activated holes; the distance between this gate and the control gate should be large to reduce the transparency of tunnel junction at the drain and get rid of ambipolar leakage.

\subsection*{Model of the interband tunneling enhanced by van Hove singularities}

Our modeling of graphene bilayer TFET relies on a self-consistent determination of the carrier density and band structure~\cite{McCann_EPJ} followed by the calculation of tunnel current under assumption of ballistic transport (see supplementary material, sections I--III). However, the principal dependence of the tunnel current on the gate voltage can be derived in a very simple fashion. The current is proportional to the number of electrons capable of tunneling between the valence band of source and the conduction band of channel. Once these band overlap by $dE$ in the energy scale, the electrons available for tunneling in graphene bilayer occupy a ring in the momentum space (Fig.~\ref{F1} B, left panel). Their number is proportional to $p_{\min}dE$, where $p_{\min}$ is the momentum corresponding to the bottom of the 'Mexican hat'. One thus concludes that the tunnel current is a linear function of the band overlap which, in turn, is a linear function of the gate voltage. This contrasts with the 2d materials having parabolic bands where the number of electrons available for tunneling is proportional to $\sqrt{E}dE$ (Fig.~\ref{F1} B, right panel). As a result, the current in TFETs based on these materials is proportional to the gate voltage raised to the power $3/2$.

A rigorous expression for the tunnel current density involves an integral of the single-particle velocity $v_{\parallel} = dE/dp_{\parallel}$ timed by the barrier transparency ${\cal D}(p_\bot, E)$ and the difference of occupation functions in the valence and conduction bands $f_v(E) - f_c(E)$ over the momentum space $d^2{\bf p} = 2 d p_{\bot} dp_{\parallel}$ ~\cite{Kane_Theory_of_tunneling}:
\begin{equation}
\label{Tun_current}
J_{\rm t} = \frac{ g_s g_v e}{h^2}\int\limits_{E_c}^{E_v}{ dE \left[f_v(E) - f_c(E)\right]} \int\limits_{0}^{p_{\max}(E)}{2dp_\bot \times 2 {\cal D} (E,p_\bot)}.
\end{equation}
Here, $g_s g_v = 4$ is the spin-valley degeneracy factor in graphene, $p_{\max}(E)$ is the maximum transverse momentum of electron at a given energy $E$, $p_{\max}(E) = \min \left\{ p_c (E), p_v (E) \right\}$, where $p_c (E)$ and $p_v (E)$ are the inverse functions to the electron dispersion in the conduction and valence bands. The limits of integration over energy are the conduction band edge in the channel, $E_c$, and the valence band edge in the source, $E_v$. The factor of two before the quasi-classical barrier transparency comes from the presence of two turning points with zero group velocity in the GBL dispersion at which an electron attempts to tunnel.

The effect of 'Mexican hat' on the current switching steepness can be traced analytically from Eq.~\ref{Tun_current} by assuming that the conduction band states are empty, valence band states are occupied, and the barrier transparency ${\cal D} (E,p_\bot)\approx{\cal D}_0$ weakly depends on the energy and transverse momentum. At small band overlap, the momenta of the tunneling electrons in graphene bilayer are close  to $p_{\min}$ (Fig. 1B, left panel), which results in
\begin{equation}
\label{Tun_current_simple}
J_{\rm t} \approx 4 g_s g_v {\cal D}_0 \frac{e p_{\min} }{h^2}(E_v - E_c).
\end{equation}
This linear dependence is in agreement with the above qualitative considerations. Previously, such a dependence of the tunnel current on the band overlap was attributed just to the 1D semiconductor structures which proved to be among the best candidates for the TFETs~\cite{Tomioka_InAsNWTFET,Jena_GNR_exp,Appenzeller}.

An additional increase in the graphene bilayer TFET subthreshold steepness occurs due to the dependence of the transparency ${\cal D} (E,p_\bot)$ on the junction field and, hence, gate voltage. The transparency is evaluated by integrating the imaginary part of the electron momentum inside the band gap, which results in (see Supporting information, section III)
\begin{equation}
\label{Tun_probability}
{\cal D} (E,p_\bot) \approx \exp\left[-\frac{\pi l}{2 \hbar} {\rm Im} p_{\parallel}(E=0)\right],
\end{equation}
where ${\rm Im} p_{\parallel}(E=0)$ is the imaginary part of electron momentum evaluated at the midgap, and $l$ is the length of the classically forbidden region (tunneling path length). The latter is given by $l = \tilde\Delta/ e F$ for $p_{\bot} < p_{\min}$, and $l = 2 E(p_{\perp})/eF$ for $p_{\bot} > p_{\min}$, where $F$ is the electric field at the junction found from the solution of Poisson's equation, and $\tilde \Delta$ is the band gap in the GBL. The thermionic leakage currents were evaluated with equations similar to (\ref{Tun_current}) by setting the unity transmission probability and constraining the energy integral to the particles with the energies above the barrier.

\begin{figure}
\centering
\includegraphics[width=.9\linewidth]{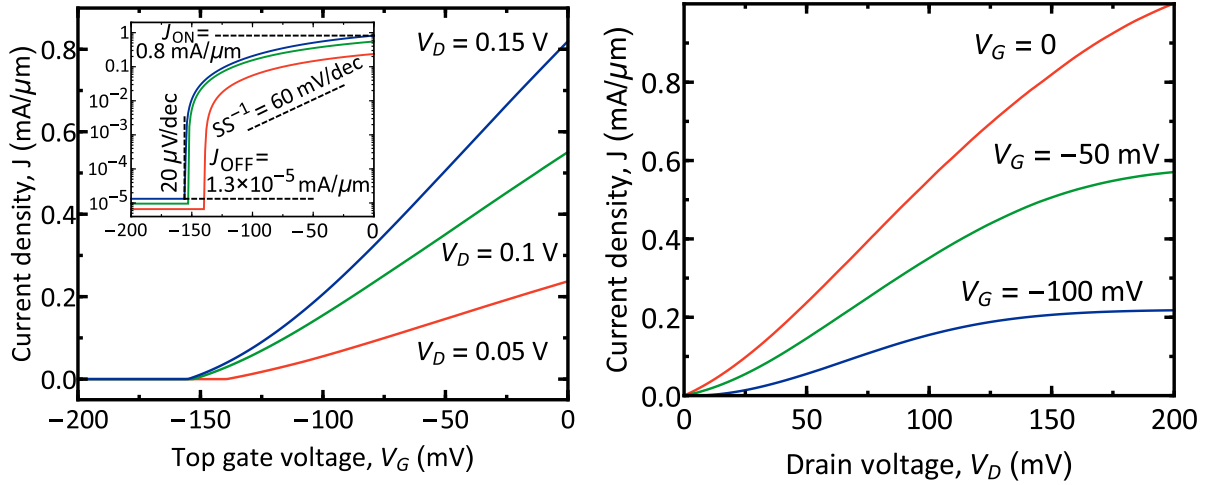}
\caption{Calculated room-temperature gate transfer (left) and current-voltage (right) characteristics of graphene bilayer TFET at fixed bias voltages at auxiliary gates: $V_B = 3.3$ V, $U_S= - 0.6$ V, $U_D = 0.25$ V. Top gate dielectric is 2 nm ZrO$_2$, $\kappa = 25$, back gate dielectric is 10 nm SiO$_2$, spacing between the gates $d_g = 5$ nm. Inset: gate transfer characteristic in the log scale}
\label{F3}
\end{figure}

\subsection*{Characteristics of the graphene bilayer TFET}

The calculated room-temperature $J(V_G)$--characteristics of graphene bilayer TFET at different drain bias $V_D$ are shown in Fig.~\ref{F3} A. The  current density just above the threshold voltage is a linear function of $V_G$, in agreement with the simple density-of-states arguments and Eq.~(\ref{Tun_current_simple}). With increasing the top gate voltage, the $J(V_G)$ - curve becomes superlinear, which is attributed to the exponential sensitivity of the tunnel barrier transparency to the junction field. The subthreshold slope at $V_G = V_{th}$ reaches ($20$ $\mu$V/dec)$^{-1}$ and is limited by the small thermionic current $J_{th}\simeq 1.3 \times 10^{-5}$~mA/$\mu$m and the gate leakage current $J_{g}\simeq 1.0 \times 10^{-5}$~mA/$\mu$m. However, it is not the highest subthreshold slope that determines the power efficiency of the TFET but rather the supply voltage $V_S$ required to switch the transistor between the ON- and OFF-states. Considering the current at $V_G = 0$ V as the ON-state current ($J_{ON} = 0.8$ mA/$\mu$m in Fig.~\ref{F3} B at $V_D = 0.15$ V), and the leakage current as the OFF-state current ($J_{ON}/J_{OFF} = 3.5 \times 10^4$), we have obtained $V_S = 150$ mV. In a conventional MOSFET, the gate voltage swing $V_S \ge 285$ mV is required to achieve the same current switching ratio.

The drain characteristics of graphene bilayer TFET shown in Fig.~\ref{F3} B demonstrate a pronounced current saturation typically absent in single graphene layer FETs. This saturation is due to the limited energy range in which the tunneling injection is possible. The presence of saturation is important for the logic inverters which guarantees the clear discrimination of the zero and unity output signals. 

The average subthreshold slope of our TFET over 4.5 decades of current is 33 (mV/dec)$^{-1}$. With this characteristic, it outperforms all sub-thermal tunnel switches~\cite{Seabaugh_State_of_Art} based on silicon~\cite{Choi}, germanium~\cite{krishnamohan}, III-V hetero junctions~\cite{Tomioka_InAsNWTFET}, and carbon nanotubes~\cite{Appenzeller} reported to date. Only recently a vertical TFET based on MoS$_2$/germanium junction with a similar value of the average subthrehold slope was demonstrated~\cite{Sarkar}, however, its ON-state current density of 1 $\mu$A/$\mu$m leaves much to be desired.

The aggregate quality of the TFET, accounting for both average subthreshold slope and current density, can be characterized by an $I_{60}$--figure of merit~\cite{Vandenberghe_APL} which is the current density at the point where the subthreshold slope equals (60 mV/dec)$^{-1}$. While the best $I_{60}$ reported to date equals 6 nA/$\mu$m (InAs nanowire/Si heterojunction TFET~\cite{Tomioka_InAsNWTFET}), in our TFET structure $I_{60}=150$~$\mu$A/$\mu$m. 

The unique characteristics of GBL TFET surpassing the existing TFETs are enabled by the three factors. First of all, it is the small extrinsic band gap (for doping gate voltages used in Fig.~\ref{F3}, $\tilde \Delta \approx 0.3$ eV) that guarantees elevated tunneling probability (${\cal D} \sim 0.1$) and large current density. Most two-dimensional transition metal dichalcogenides have large intrinsic band gaps ($1.9$ eV for MoS$_2$, 1.3 eV for WS$_2$, etc.), while in the 2D structures based on III-V materials being narrow-gap in the bulk, the gap value increases significantly due to the quantum confinement~\cite{Jena_IEEE_TFETSon2D}. Secondly, the singular DoS near the band edges allows an abrupt switching of tunnel current. Even if there existed a parabolic-band 2D material with the same band gap and the same barrier transparency in the TFET structure, its current density would be given by (see supplemental material, section IV)
\begin{equation}
\label{Tun_current_parabolic}
J_{\rm {t, par}} \approx \frac{4g_s}{3}\frac{e}{h^2} {\cal D}_0 \sqrt{\frac{2m_c m_v}{m_c + m_v}} (E_v - E_c)^{3/2},
\end{equation}
\begin{figure}
\centering
\includegraphics[width=.6\linewidth]{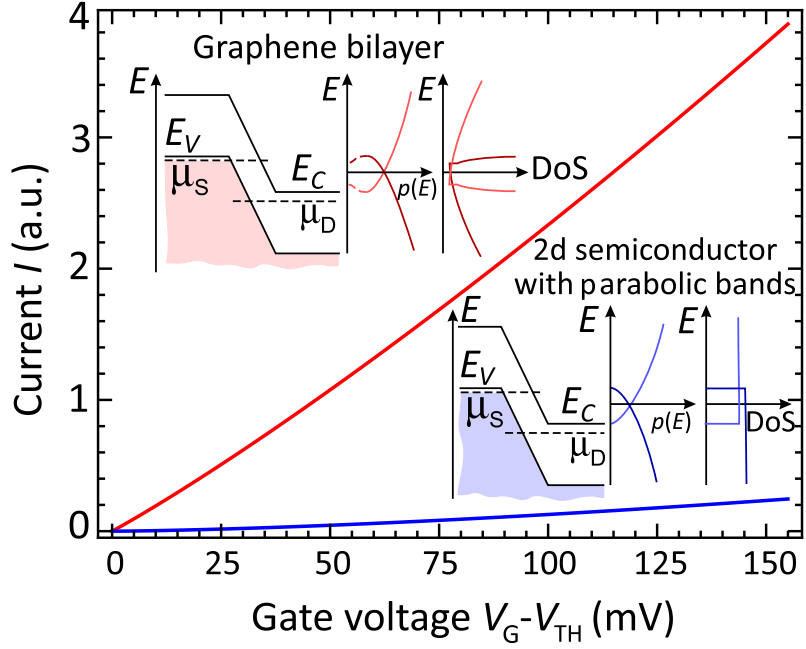}
\caption{Comparison of the gate transfer characteristics of GBL TFET and a TFET based on an equivalent 2D semiconductor with the same barrier transparency ${\cal D}_0$, but with different (parabolic) band structure. Numerical values of the effective masses are taken for bulk InAs. The insets show the band diagrams overlaid with electron-hole spectra and the energy dependence of DoS
}
\label{F5}
\end{figure}
where $m_c$ and $m_v$ are the conduction and valence band effective masses and, similar to the derivation of Eq.~(\ref{Tun_current_simple}), we have assumed the barrier transparency ${\cal D}_0$ to be energy- and momentum independent. The numerical comparison of current density in graphene bilayer and its equivalent parabolic band counterpart is presented in Fig. 4 for the effective mass values typical for narrow-gap III-V semiconductors ($m_c = 0.024 m_0$, $m_v = 0.026 m_0$ for InAs). At 150 mV gate voltage above the threshold, the current density in graphene bilayer exceeds 15 times that in a parabolic-band material. The factor of two is due to the valley degeneracy absent in III-V's, another factor of two is due to the tunneling at two turning points of the 'Mexican hat' dispersion, and the remainder of $3.5$ is due to the finiteness of electron momentum at the edge of the 'Mexican hat'.

\subsection*{Gate leakage and band tailing: the insulator selection rules}

The steep switching of the tunnel current by the gate voltage can be masked by the leakage to the gates, band-tail and trap-assisted tunneling~\cite{Vandooren_SSE,Trap-Assisted,Khayer_JAP}. The latter factors might have masked the onset of the interband current in the recent measurements of graphene bilayer tunnel junctions~\cite{Miyazaki_Tunnel,Miyazaki_Nanoscale}. A careful selection of the gate dielectrics providing high interface quality is required to minimize these effects.

%These dopants also form discrete energy levels inside the band gap, providing additional pathways for the electron tunneling in the OFF-state. 

\begin{figure}
\centering
\includegraphics[width=.6\linewidth]{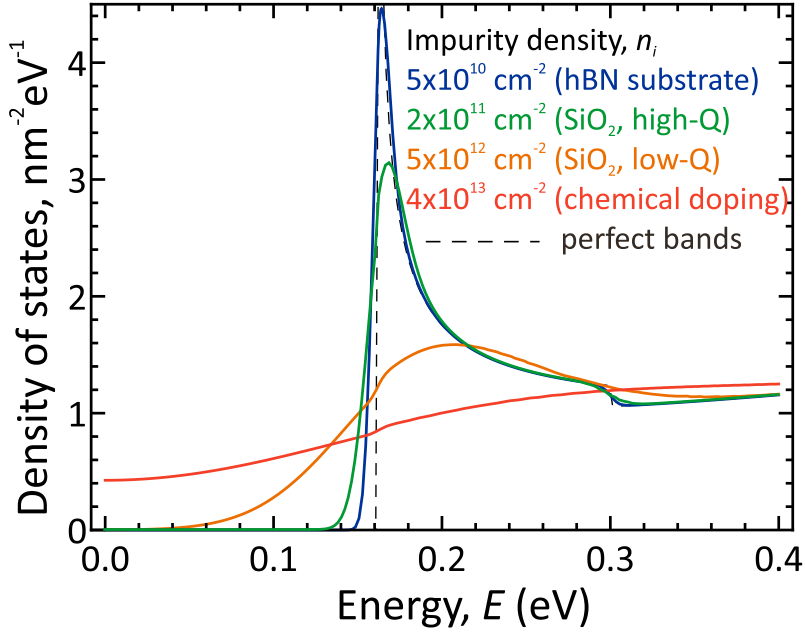}
\caption{Calculated energy dependencies of the DoS in the conduction band of graphene bilayer at different densities of charged impurities (corresponding to the substrates of different quality). The electron density is held fixed at $4 \times 10^{13}$ cm$^{-2}$, the nominal energy gap is 0.3 eV}
\label{DoS}
\end{figure}

The main reason for the band tailing comes from the fluctuations of electric potential produced by the random charged defects or dopants~\cite{Kane_Tails}. This effect is most pronounced in the TFETs with source and drain intentionally doped chemically. In the TFETs with electrically doped contacts, only residual charged impurities inevitably present in the substrate contribute to the band tailing. To provide a quantitative view on the band tailing in graphene bilayer on different substrates, we have evaluated the quasi-classical DoS $\rho(E)$ in the presence of fluctuating potential by integrating the singular 'bare' DoS $\rho_0$ over the probabilities of voltage fluctuations~\cite{Kane_Tails}
\begin{equation}
\rho(E) = \frac{1}{\sqrt{2\pi \langle V^2\rangle}} \int\limits_{-\infty}^{+\infty}{\rho_0(E - e V) \exp\left(-\frac{V^2}{2\langle V^2\rangle} \right) dV },
\end{equation}
where $\langle V^2\rangle$ is the root-mean-square amplitude of the voltage fluctuations proportional to the impurity density $n_i$. The calculated energy dependencies of the 'smeared' DoS are shown in Fig.~\ref{DoS}. For the parameters of chemical doping used in the pioneering proposal of the GBL TFET, $n_i = 4\times 10^{13}$ cm$^{-2}$, the conduction and valence bands almost merge together, which would result in a poor OFF-state, nothing to say about high switching steepness. A slight peak in the DoS near the band bottom becomes noticeable already at impurity density of $5 \times 10^{12}$ cm$^{-2}$ which corresponds to the low-quality graphene on SiO$_2$ substrates. In graphene samples on a high-quality SiO$_2$~\cite{Tan_GrapheneSiO2}, the smearing of the band edge is order of $10$ meV. The ultimate band abruptness of $\sim 5$ meV can be achieved in graphene samples encapsulated in boron nitride, providing the residual impurity density of $\sim 5 \times 10^{10}$ cm$^{-2}$~\cite{GBL_encapsulated_in_BN}. At this limit, the fluctuation-induced smearing of the bands becomes negligible, and the behavior of the DoS near the bottom of the ''Mexican hat'' is governed by the trigonal warping distortions of electron spectrum due to the next-nearest neighbor interactions~\cite{McCann_EPJ}. Using the exact spectrum of GBL with trigonal warping, we estimate the energy scale where the trigonal warping is relevant as $\delta\varepsilon \approx 20$ meV. Already for relatively small gate voltages, $V_G - V_{th} > \delta \varepsilon/e$, these corrections are irrelevant and the linearity of the $J(V_G)$--characteristic holds.

The gate leakage may also limit the minimum achievable OFF-state current, while at the same time small effective gate oxide thickness is required to efficiently control the band structure in the channel by the gate voltage. Among the common high-$\kappa$ materials, zirconium oxide ($\kappa \approx 25$) looks as an optimal solution for the GBL TEFT due to the large band offset with respect to graphene ($U_b = 2.9$ eV~\cite{ZrO2bandoffset}) and elevated tunneling mass $m_t \approx 0.3 m_0$~\cite{ZrO2mass1}. We have evaluated the leakage current from graphene with electron (hole) density of $n_{e(h)}$ into the metal gate to be (see Supporting information, section VI)
\begin{equation}
J_g = 8 e L_g n_{e(h)} \frac{U_b}{\hbar}\frac{k_F l_{\rm loc}}{1+(k_F l_{\rm loc})^2}{\cal D}_g,
\end{equation} 
where $l_{\rm loc} = \hbar/\sqrt{2m_t U_b}$ is the electron localization length in the direction perpendicular to the graphene bilayer, $k_F$ is the Fermi wave vector in the metal, ${\cal D}_g$ is the transparency of the barrier separating GBL and the gate, and $L_G$ is the gate length. Under the biasing conditions of Fig.~\ref{F3}, the gate leakage current is estimated to be $J_t = 1.0 \times 10^{-5}$ mA/$\mu$m which is below the thermionic leakage level (in this estimate, we have taken $L_g=20$ nm and $k_F = 2$ \AA$^{-1}$).

\section*{Discussion}

We have proposed and substantiated the operation of a graphene bilayer TFET exploiting the van Hove singularities in the density of states near the band edges. The presence of these singularities leads to the increased steepness of the gate characteristics and to the high ON-state current as well. The subthreshold slope of $J(V_G)$ curve in the proposed FET reaches the maximum of (20 $\mu$V/dec)$^{-1}$, while only 150 mV gate voltage swing is required to change the current density from $J_{ON} \approx 1$ mA/$\mu$m down to  $J_{OFF} \approx 2 \times 10^{-5}$ mA/$\mu$m. As a matter of fact, the effects of singular DoS on the interband tunneling are possible just in the TFET structures with an all-electrical doping, where the effects of band tailing and trap-assisted tunneling are minimized.

Such steep switching in the lateral TFETs based on 2d materials is possible if only the van Hove singularities are present both at the top of the valence and the bottom of the conduction band. This property is unique to the graphene bilayer and is absent in other 2d materials (e.g., those based on III-V compounds), where a 'Mexican hat' structure is formed in one of the bands due to the spin-orbit coupling~\cite{Ganichev_PSS}. We can thus conclude that graphene bilayer is an only two-dimensional material where the switching of interband tunnel current is as steep as in one-dimensional semiconductors, whereas the large on-state current is inherited from the single layer graphene. 

\section*{Methods}
The modeling of GBL TFET is based on the self-consistent determination of carrier density and band structure under fixed gate voltages~\cite{McCann_EPJ} followed by the calculation of tunnel current with Eq.~(\ref{Tun_current}). The necessity for self-consistent calculation is dictated by the dependence of the energy gap on the electric field between graphene layers comprising the GBL; the field, in turn, depends on the induced carrier density which is sensitive to the band structure.  The distribution of electric field at the tunneling junction required for the evaluation of the barrier transparency is calculated with the conformal mapping technique. The numerical model is described in detail in Supporting information, sections I-II. In section III of the Supporting information, an approximate analytic model of GBL TFET is presented. 

The effect of charged impurities present in the substrate on the singular density of states in graphene bilayer is evaluated with Kane's quasi-classical model of band tails~\cite{Kane_Tails}. The revision of the model for the two-dimensional GBL is presented in Supporting information, section V.

The gate leakage current is estimated with a quantum-mechanical model of graphene bilayer tunnel coupled to the continuum of delocalized states in the metal gate. The model is presented in Supporting information, section VI.

\section*{Acknowledgements}
The work was supported by the Russian Scientific Foundation (Project \# 14-29-00277) and by the Russian Foundation for Basic Research (Grant \# 14-07-00937). The work at RIEC was supported  by the Japan Society for Promotion of Science (Grant-in-Aid for Specially Promoted Research \# 23000008).

\section*{Author contributions statement}
D.S. conceived the idea of tunnel current switching enhanced by the van Hove singularities in graphene bilayer, G.A. developed the model of interband tunneling and calculated the transistor characteristics. V.V. and V.R. proposed the device structure. G.A. and D.S. wrote the manuscript. All authors reviewed the manuscript. 

\section*{Additional information}
\subsection*{Competing Financial Interests statement}
The authors declare no competing financial interests.

\newpage

\title{Supporting information to ''Abrupt current switching in graphene bilayer tunnel transistors enabled by van Hove singularities''}

\author[1,2]{Georgy Alymov}
\author[1,2]{Vladimir Vyurkov}
\author[3]{Victor Ryzhii}
\author[1,2,*]{Dmitry Svintsov}

\affil[1]{Department of Physical and Quantum Electronics, Moscow Institute of Physics and Technology, Dolgoprudny 141700, Russia}
\affil[2]{Laboratory of Sub-micron Devices, Institute of Physics and Technology  RAS, Moscow 117218, Russia}
\affil[3]{Research Institute of Electrical Communication, Tohoku University, Sendai 980-8577, Japan}

\affil[*]{svintcov.da@mipt.ru}

%\keywords{Keyword1, Keyword2, Keyword3}

\begin{abstract}
In a continuous search for the energy-efficient electronic switches, a great attention is focused on tunnel field-effect transistors (TFETs) demonstrating an abrupt dependence of the source-drain current on the gate voltage. Among all TFETs, those based on one-dimensional (1D) semiconductors exhibit the steepest current switching due to the singular density of states near the band edges, though the current in 1D structures is pretty low. In this paper, we propose a TFET based on 2D graphene bilayer which demonstrates a record steep subthreshold slope enabled by van Hove singularities in the density of states near the edges of conduction and valence bands. Our simulations show the accessibility of $3.5\times10^4$ ON/OFF current ratio with 150 mV gate voltage swing, and a maximum subthreshold slope of (20 $\mu$V/dec)$^{-1}$ just above the threshold. The high ON-state current of  $0.8$~mA/$\mu$m is enabled by a narrow ($\sim 0.3$ eV) extrinsic band gap, while the smallness of the leakage current is due to an all-electrical doping of the source and drain contacts which suppresses the band-tailing and trap-assisted tunneling.
\end{abstract}

\flushbottom
\maketitle
\thispagestyle{empty}

\section{Self-consistent evaluation of band gap and carrier density in gated bilayer}

The intrinsic graphene bilayer (GBL) is a gapless semiconductor with symmetric electron-hole dispersion being parabolic near the band edges. However, the application of transverse electric field $F_{\perp}$ leads to the potential energy difference $\Delta = e F_{\perp} d$ between the layers comprising GBL ($d = 3.35$ A is the interlayer separation in GBL) and opening of the band gap. As the transverse electric field depends on the carrier density, which, in turn, depends on the band structure, a self-consistent procedure for the determination of those quantities in gated GBL is required.

\begin{figure}[t]
\centering
\includegraphics[width=0.5\linewidth]{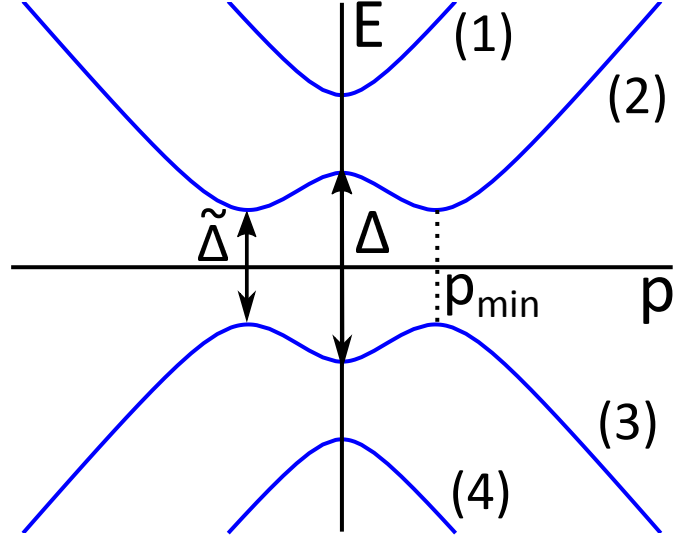}
\caption{Band structure of graphene bilayer in tight binding approximation. $\Delta$ is the interlayer asymmetry parameter, $\tilde \Delta = \frac{\Delta \gamma_1}{\sqrt{\Delta^2+\gamma_1^2}}$ is the bandgap, and $p_{min}=\frac{\sqrt{\Delta^2 + \tilde \Delta^2}}{2v}$ is the electron momentum at the band edges \label{bands} }
\end{figure}

In this section, we describe the procedure used to evaluate the carrier density, bandgap, and electric potential in GBL at given potentials of top and bottom gates ($V_t$ and $V_b$). The tight-binding Hamiltonian of graphene bilayer
\begin{equation}
\label{hamiltonian}
\hat H\left( {\bf p} \right) = \left( {\begin{array}{*{20}c}
   { \frac{\Delta }{2}} & {v(p_x  - ip_y )} & {\gamma _1 } & 0  \\
   {v(p_x  + ip_y )} & { \frac{\Delta }{2}} & 0 & 0  \\
   {\gamma _1 } & 0 & {-\frac{\Delta }{2}} & {v(p_x  + ip_y )}  \\
   0 & 0 & {v(p_x  - ip_y )} & {-\frac{\Delta }{2}}  \\
\end{array}} \right)
\end{equation}
yields the following energy bands depicted in Fig.~\ref{bands}~\cite{McCann_ElProperties}:

\begin{gather}
\label{spectrum}
E_{1,4} \left( {\bf p} \right) =  \pm \sqrt {\frac{{\gamma _1^2 }}{2} + \frac{{\Delta ^2 }}{4} + v^2 p^2  + \sqrt {\frac{{\gamma _1^4 }}{4} + v^2 p^2 \left( {\gamma _1^2  + \Delta ^2 } \right)} } , \\ 
 E_{2,3} \left( {\bf p} \right) =  \pm \sqrt {\frac{{\gamma _1^2 }}{2} + \frac{{\Delta ^2 }}{4} + v^2 p^2  - \sqrt {\frac{{\gamma _1^4 }}{4} + v^2 p^2 \left( {\gamma _1^2  + \Delta ^2 } \right)} } ,
\end{gather}

with the corresponding four--component eigen wave functions (non-normalized):
\begin{equation} \label{wavefunction}
\Psi _i \left( {\bf p} \right) = \left( {\begin{array}{*{20}c}
   {\Psi _i^{A_{top} } \left( {\bf p} \right)}  \\
   {\Psi _i^{B_{top} } \left( {\bf p} \right)}  \\
   {\Psi _i^{A_{bottom} } \left( {\bf p} \right)}  \\
   {\Psi _i^{B_{bottom} } \left( {\bf p} \right)}  \\
\end{array}} \right) = \left( {\begin{array}{*{20}c}
   { \left[ {\frac{\Delta }{2} - E_i \left( {\bf p} \right)} \right]\left\{ {\left[ {\frac{\Delta }{2} + E_i \left( {\bf p} \right)} \right]^2  - v^2 p^2 } \right\}}  \\
   {-v\left( {p_x  + ip_y } \right)\left\{ {\left[ {\frac{\Delta }{2} + E_i \left( {\bf p} \right)} \right]^2  - v^2 p^2 } \right\}}  \\
   {\gamma _1 \left[ {\frac{{\Delta ^2 }}{4} - E_i^2 \left( {\bf p} \right)} \right]}  \\
   {\gamma _1 v\left( {p_x  - ip_y } \right)\left[ {\frac{\Delta }{2} - E_i \left( {\bf p} \right)} \right]}  \\
\end{array}} \right).
\end{equation}
The components of wave function determine the amplitudes of finding an electron on the atoms of type ''A'' and ''B'' of the unit cell on the top and bottom layers in the $i$-th band. In Eqs. (\ref{hamiltonian}) -- (\ref{wavefunction}), ${\bf p}$ is the electron quasimomentum, ${\gamma _1 \approx}$ 0.4 eV is the interlayer hopping integral, and $v \approx 10^6$ m/s is the Fermi velocity in graphene. Given the eigenfunctions, the probabilities of finding an electron of the $i$-th energy band on the top and the bottom graphene layer can be calculated as follows:
\begin{gather}
\label{probabilities}
 w_i^{t} \left( {\bf p} \right) = \frac{{\left| {\Psi _i^{A_{top} } \left( {\bf p} \right)} \right|^2  + \left| {\Psi _i^{B_{top} } \left( {\bf p} \right)} \right|^2 }}{{\left\| {\Psi _i \left( {\bf p} \right)} \right\|^2 }}, \\ 
 w_i^{b} \left( {\bf p} \right) = \frac{{\left| {\Psi _i^{A_{bottom} } \left( {\bf p} \right)} \right|^2  + \left| {\Psi _i^{B_{bottom} } \left( {\bf p} \right)} \right|^2 }}{{\left\| {\Psi _i \left( {\bf p} \right)} \right\|^2 }};
\end{gather}
here $||...||$ stands for the norm of the vector.

The net charge densities at top and bottom graphene layers are obtained by summing the probabilities (\ref{probabilities}) timed by the distribution functions over the bands and momenta:
\begin{gather}
\label{top density}
\rho _t  = eg \left\{ {\sum\limits_{{\bf p},i=3,4} {w_i^t \left( {\bf p} \right)
\left[ {1 - f\left( {E_i \left( {\bf p} \right),\mu } \right)} \right]} }  -  {\sum\limits_{{\bf p}, i=1,2} {w_i^t \left( {\bf p} \right)f\left( {E_i \left( {\bf p} \right),\mu } \right)} } \right\}, \\
\label{bottom density}
\rho _b  = eg\left\{ {\sum\limits_{{\bf p},i=3,4} {w_i^b \left( {\bf p} \right)
\left[ {1 - f\left( {E_i \left( {\bf p} \right),\mu } \right)} \right]} }  - {\sum\limits_{{\bf p},i=1,2} {w_i^b \left( {\bf p} \right)f\left( {E_i \left( {\bf p} \right),\mu } \right)} } \right\}.
 \end{gather}
Here, $g=4$ is the spin-valley degeneracy factor, $f\left( {E,\mu } \right) = [e^{\frac{{E - \mu }}{{kT}}}  + 1]^{-1}$ is the Fermi-Dirac distribution function, $\mu$ is the chemical potential reckoned from the midgap, $k$ is the Boltzmann constant, and $T$ is the absolute temperature.

\begin{figure}[t]
\centering
\includegraphics[width=0.9\linewidth]{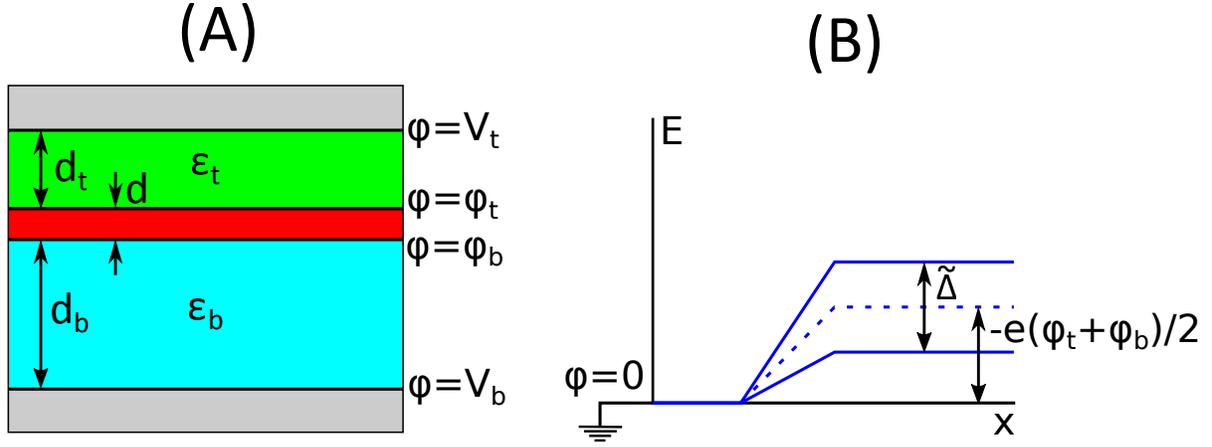}
\caption{ (A) Illustration of the electrostatic calculations. Electric field produced by top and bottom gates with potentials $V_t$ and $V_b$ induces potentials $\varphi_t$ and $\varphi_b$ at top and bottom layers of GBL. Panel (B) shows the band diagram in the source region. Near the grounded source contact there is no bandgap and GBL has zero potential, while under the "doping" gate GBL exhibits a bandgap $\tilde \Delta$ and has potential $(\varphi_t+\varphi_b)/2$, which leads to a corresponding midgap shift  \label{electrostatics} }
\end{figure}

We further consider graphene bilayer placed between the top and bottom gates separated by gate dielectrics of thicknesses $d_t$ and $d_b$ and dielectric constants $\varepsilon _t$ and $\varepsilon _b$, respectively (see Fig.~\ref{electrostatics}).

With the aid of Gauss law, the potential energy difference between layers, $\Delta = e (\varphi_b - \varphi_t)$, is readily expressed via the gate potentials $V_t$ and $V_b$:
\begin{equation} \label{Delta}
\Delta   = ed\frac{{V_b  - V_t  + \frac{{\rho _b d_b }}{{\varepsilon _0 \varepsilon _b }} - \frac{{\rho _t d_t }}{\varepsilon _0 \varepsilon _t }}}{d + d_t/\varepsilon _t + d_b/\varepsilon _b},
\end{equation}
where $\varepsilon _0$ is the vacuum permittivity. Assuming that far from gates graphene is undoped ($\mu=0$), and the source contact is grounded, we obtain an expression for  the chemical potential in the source region:
\begin{equation} \label{mu}
\mu  = e\frac{{\varphi _t  + \varphi _b }}{2} = e\frac{{\left( {V_t  + \frac{{\rho _t d_t }}{{\varepsilon _0 \varepsilon _t }}} \right)\left( {\frac{d}{2} + \frac{{d_b }}{{\varepsilon _b }}} \right) + \left( {V_b  + \frac{{\rho _b d_b }}{{\varepsilon _0 \varepsilon _b }}} \right)\left( {\frac{d}{2} + \frac{{d_t }}{{\varepsilon _t }}} \right)}}{d + d_t/\varepsilon _t + d_b/\varepsilon _b}.
\end{equation}
Equations~(\ref{spectrum})--(\ref{mu}) form a system of equations used for a self-consistent evaluation of $\Delta$ and $\mu$ in the source region. The system, which can be written in the symbolic form
\begin{equation}
\left( {\begin{array}{*{20}c}
   \Delta   \\
   \mu   \\
\end{array}} \right) = {\bf F}\left( {\Delta ,\mu } \right),
\end{equation}
was solved numerically using an iterative method similar to the nonlinear successive over relaxation~\cite{SOR}:
\begin{equation}
\left( {\begin{array}{*{20}c}
   {\Delta ^{i + 1} }  \\
   {\mu ^{i + 1} }  \\
\end{array}} \right) = \left( {1 - \lambda } \right)\left( {\begin{array}{*{20}c}
   {\Delta ^i }  \\
   {\mu ^i }  \\
\end{array}} \right) + \lambda {\bf F}\left( {\Delta ^i ,\mu ^i } \right),
\end{equation}
where $i$ is the iteration number, and $\lambda$ is the relaxation parameter adjusted to achieve convergence.
The same approach was applied for the evaluation of $\Delta$ and $\mu$ in the drain region (with an obvious shift of electric potentials by the drain voltage). Considering the GBL band structure under the control (middle) gate, we assumed that the conduction band electrons are in equilibrium with the drain contact, while the holes are in equilibrium with the source contact. This approach is justified as far as the tunneling probability is small and the electrons injected from the valence band do not distort the potential distribution. To this end, two different chemical potentials for electrons and holes were used in Eqs.~(\ref{top density}),~(\ref{bottom density}) when calculating the charge density under the control gate.

\section{Distribution of electric potential, barrier transparency, and current}

The calculation of the tunnel current requires the knowledge of the barrier transparency which, in turn, depends on the distribution of electric potential  in the tunneling region. Various studies of the tunnel diodes confirm that it is the maximum field in the junction region that governs the barrier transparency~\cite{Diode-1,Diode-2,Diode-3}. Thus, for an estimate of the tunnel current, the full knowledge of potential distribution is not necessary.

\begin{figure}[t]
\centering
\includegraphics[width=0.5\linewidth]{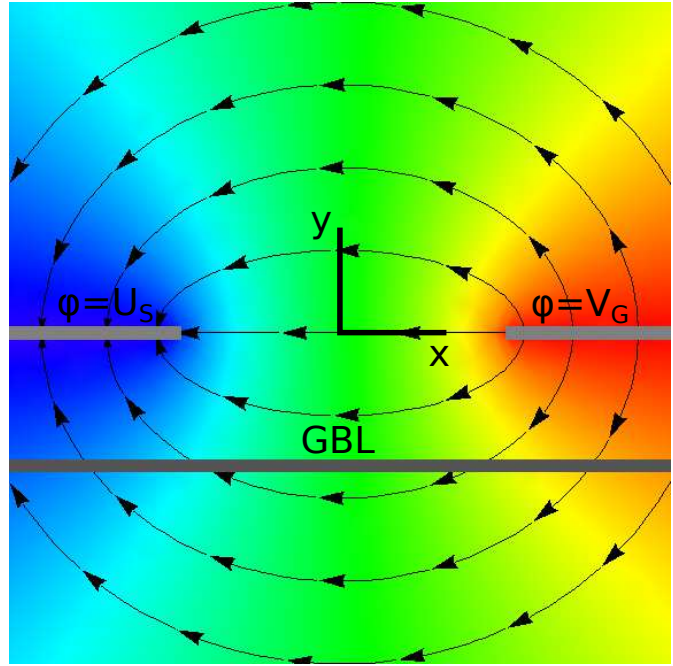}
\caption{Distribution of the electric potential produced by two keen planes held at potentials $U_S$ and $V_G$. Potential is encoded by colour, increasing from blue to red, while arrows represent field lines \label{field} }
\end{figure}

To obtain the junction field, we have solved the Laplace equation in the region between two gate electrodes: the ''doping gate'' above the source (held at potential $U_S$) and the top control gate above the channel (held at potential $V_G$) treating them as two keen metal plates~\cite{Ryzhii_Khrenov}. The lateral distance between the gates is denoted by $d_g$. The effect of the bottom gate on the junction field is weak provided $d_b \gg d_t$. The difference between dielectric constants of top and bottom dielectric layers, $\varepsilon_t \neq \varepsilon_b$, can be approximately taken into account by replacing the top gate oxide thickness with effective oxide thickness, $d_{eff} = d_t \varepsilon_b/\varepsilon_t$. The distribution of electric potential in the geometry considered is obtained with conformal mapping technique, and the result is 
\begin{equation}
\varphi(x,y) = \frac{U_S + V_G}{2} + \frac{2}{\pi}\frac{V_G-U_S}{2}{\rm Re}\left[ \arcsin \left(\frac{x + i y}{d_g/2}\right)\right]
\end{equation}
(see Fig.~\ref{field}), where the origin of the coordinate grid is located between the two gates and lies in their plane.

The maximum electric field in the graphene plane ($y = - d_{eff}$) is achieved at $x=0$ and equals:
\begin{equation}
F = \frac{{V_G  - U_S }}
{{\pi \sqrt {\left( {\frac{{d_g }}
{2}} \right)^2  + \left( {\frac{{\varepsilon _b }}
{{\varepsilon _t }}d_t } \right)^2 } }},
\end{equation}

Having obtained the junction field $F$, we calculate the WKB transparency of the potential barrier ${\cal D} \left( {p_\perp ,E} \right)$ separating the conduction and valence bands. Assuming that the potential energy depends on $x$ - coordinate linearly, we find
\begin{equation}
\label{transparency}
{\cal D} \left( {p_\perp ,E} \right) = 
\exp\left[ - \frac{2}{\hbar}\int\limits_{x_1}^{x_2} {\rm Im} p_\parallel \left( x \right)dx \right]  
= \exp\left\{ - \frac{2}{\hbar }\int\limits_{x_1}^{x_2} { {\rm Im} \sqrt {p^2 \left[ {E+eFx,\:\Delta \left( x \right)} \right] - p_{\perp}^2 } dx} \right\},
\end{equation}
where $p_\parallel$ and $p_\perp$ are the components of electron quasimomentum that are parallel and perpendicular to the direction of junction field, respectively, $E$ is the electron energy measured from the midgap in the source region, the integrals are taken between the classical turning points $x_1$ and $x_2$. Generally, the interlayer asymmetry parameters in the source ($\Delta_S$) and in the channel ($\Delta_C$) are different. For simplicity, we perform a linear interpolation  between $\Delta_S$ and $\Delta_C$ in the junction region,
\begin{equation}
\Delta \left( x \right) = \Delta _S  + \frac{{\Delta _C  - \Delta _S }}
{l}x = \Delta _S  + \frac{{\Delta _C  - \Delta _S }}
{{\varphi _C  - \varphi _S }}Fx,
\end{equation}
where $\varphi _S$ and $\varphi _C$ are the electric potentials in the source and channel regions (obtained in the previous section), and $l$ is the length of the transition region between them.

\begin{figure}[t]
\centering
\includegraphics[width=0.6\linewidth]{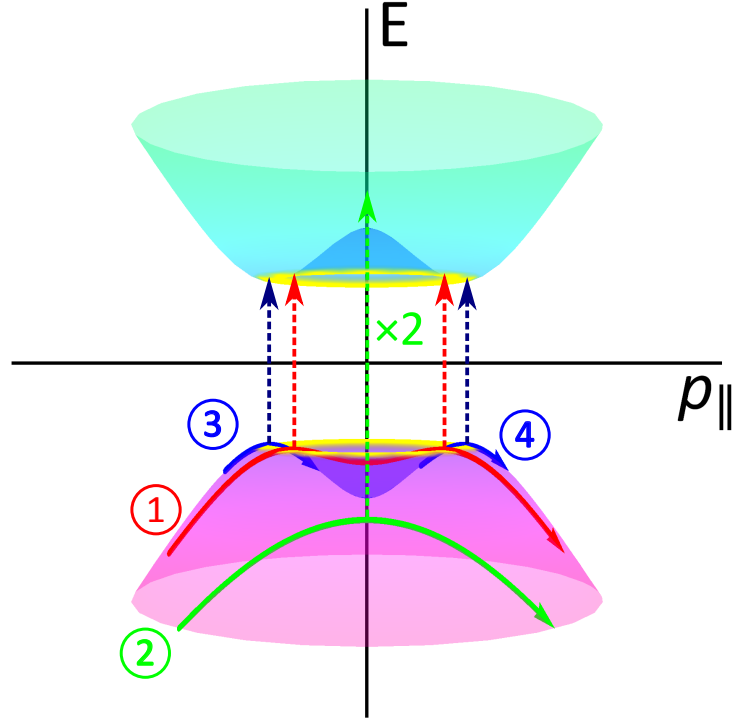}
\caption{\label{Tunneling} Illustration of interband tunneling in bilayer graphene. Valence band electrons with small transverse momentum and energy far from the band edges (1) pass through the two extrema of the dispersion law while moving in the constant electric field and may tunnel from each of these extrema. Electrons with large transverse momentum (2) pass through just one maximum. However, during their motion in the forbidden region, there are also two tunneling paths. Finally, for the energies close to the valence band edge in Eq.~(\ref{tunnelcurrent}), $ - \frac{\Delta}{2} < E < - \frac{\tilde \Delta}{2}$ , there are two electron states with positive group velocity [(3) and (4)] which contribute to the current independently}
\end{figure}

The tunneling current density $J_{tun}$ can be written as
\begin{equation}
\label{tun}
J_{tun}  = eg\sum\limits_{\substack{
{\bf p}\\
v_\parallel>0}}{v_\parallel\left( {\bf p} \right){\cal D}\left( {\bf p} \right) \left[ {f_S \left( {\bf p} \right) - f_C \left( {\bf p} \right)} \right]}, 
\end{equation}
where $v_\parallel= {\partial E}/{\partial p_\parallel}$ is the component of electron group velocity parallel to the current direction, and $f_S$ and $f_C$ are the Fermi-Dirac distributions in the source and channel, respectively. It is more convenient to express all quantities in Eq.~(\ref{tun}) via the {\it transverse} electron momentum $p_\bot$ and its total energy $E$ which are conserved during the tunneling:
\begin{equation}
\label{tunnelcurrent}
J_{tun}  = 2eg\int\limits_{E_c }^{E_v } {\left[ {f_S \left( E \right) - f_D \left( E \right)} \right]\frac{{dE}}{{2\pi \hbar }}\int\limits_{ - p_{\max } \left( E \right)}^{ + p_{\max } \left( E \right)} {{\cal D}\left( {p_{\perp},E} \right)\frac{{dp_{\perp}}}{{2\pi \hbar }}} }, 
\end{equation}
where $f_S \left( E \right) \equiv f\left( {E,\mu _S } \right)$, $f_D \left( E \right) \equiv f\left( {E,\mu _D } \right)$, $\mu _S$ and $\mu _D$ are the chemical potentials in source and drain regions, respectively (we assume the electrons in the conduction band of the channel region are in equilibrium with the drain contact), $p_{\max } \left( E \right)$ is the maximum transverse momentum allowed for given total electron energy $E$ in the valence band of source and in the conduction band of the channel. More precisely, $p_{\max}(E) = \min\left\{ p_c(E), p_v(E) \right\}$, where $p_c(E)$ and $p_v(E)$ are the inverse functions of the electron dispersion law $E(p)$ in the conduction band of the channel and the valence band of source, respectively~\cite{Kane_Theory_of_tunneling}. Finally, we note the appearance of the factor of two before the whole expression for the tunnel current which is unique to the band structure of graphene bilayer. It appears due to the presence of two points with zero group velocity in the dispersion laws for electrons and holes, which implies that a carrier incident on a barrier attempts to pass through it twice (see the red arrow in Fig.~\ref{Tunneling}).

Thermionic currents $J_e$ and $J_h$ carried by electrons and holes, respectively, are calculated in a similar way, except for the absence of the factor of two and barrier transparency equal to unity. The integration limits $E$ in the expressions for the thermionic current are restricted to the energies above the barriers created by the doping gates:
\begin{gather}
\label{thermionic}
J_e  = \frac{{eg}}{{2\pi ^2 \hbar ^2 }}\int\limits_{E_c^S }^{ + \infty } {p_{\max } \left( E \right)\left[ {f_S \left( E \right) - f_D \left( E \right)} \right]dE}, \\
J_h  = \frac{{eg}}{{2\pi ^2 \hbar ^2 }}\int\limits_{ - \infty }^{E_v^D } {p_{\max } \left( E \right)\left[ {f_S \left( E \right) - f_D \left( E \right)} \right]dE}, 
\end{gather}
where $E_c^S$ and $E_v^D$ are the edges of conduction band in source region and valence band in drain region, respectively. Electrons from the "hat" (i.e. with $p<p_{min}$) do not contribute to the thermionic current because they can neither remain in the "hat" in all three regions (source, gate and drain) nor get out of it (it is possible only at band edges).
\section{Derivation of approximate analytic expressions for tunnel and thermionic currents}
\begin{figure}[t]
\centering
\includegraphics[width=0.6\linewidth]{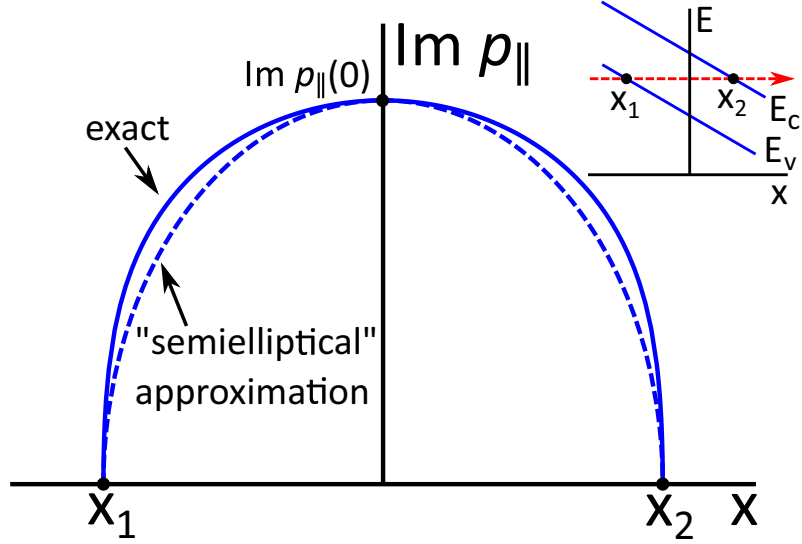}
\caption{Exact and approximate imaginary part of electron longitudinal momentum in the barrier. Electron with given energy and transverse momentum exits valence band at point $x_1$, tunnels through the bandgap, and enters conduction band at point $x_2$, as shown in the inset \label{elliptic} }
\end{figure}

Firstly, we need to obtain an analytic expression for barrier transparency. The integrand in Eq.~(\ref{transparency}) is too complicated to carry out the integration exactly. However, using the fact that the integrand has an arch-like shape (it turns to zero at $x_1$ and $x_2$ and reaches maximum somewhere between), we can approximate it with a half ellipse of the same width and height (see Fig.~\ref{elliptic}).

In the simplest case of constant $\Delta=\Delta_{tun}$ the maximum of the integrand is reached in the middle of the barrier and equals
\begin{equation}
\label{barrierheight}
\max {\mathop{\rm Im}\nolimits}\, p_{\parallel}\left( x \right) = {\mathop{\rm Im}\nolimits} \sqrt {p^2 \left( {0,\Delta _{tun} } \right) - p_{\perp}^2 }  = {\mathop{\rm Im}\nolimits} \sqrt {\frac{1}{v^2}\left(\frac{{\Delta _{tun}^2 }}{4} + \frac{{\gamma _1 \Delta _{tun} }}{2}i\right) - p_{\perp}^2 }. 
\end{equation}
The main contribution to the tunnel current is from electrons with $p_{\perp}<p_{min}$, because for them barrier is the lowest and the shortest. Such electrons tunnel from the edge of the valence band to the edge of the conduction band, which gives the following barrier length:
\begin{equation}
\label{barrierlength}
x_2  - x_1  = \frac{{\tilde \Delta _{tun} }}{{eF}}.
\end{equation}
Equations (\ref{transparency}), (\ref{barrierheight}) and (\ref{barrierlength}) yield the following expression for barrier transparency in the "semielliptical" approximation:
\begin{equation}
\label{semielliptical}
{\cal D}\left( p_{\perp} \right) \approx \exp \left( { - \frac{{\pi \tilde \Delta _{tun} }}{{2\hbar eFv}}\,{\mathop{\rm Im}\nolimits} \sqrt {\frac{{\Delta _{tun}^2 }}{4} + \frac{{\gamma _1 \Delta _{tun} }}{2}i - v^2 p_{\perp}^2 } } \right),
\end{equation}
which does not depend on electron energy.
To be able to perform the integration with respect to $p_{\perp}$ in Eq.~(\ref{tunnelcurrent}), we expand the exponent in Eq.~(\ref{semielliptical}) in a Maclaurin series up to the second order:
\begin{equation}
{\cal D}\left( p_{\perp} \right) \approx {\cal D}_0 \exp \left( { - \frac{{p_{\perp}^2 }}{{p_0^2 }}} \right),
\end{equation}
where
\begin{equation}
{\cal D}_0  = \exp \left( { - \frac{{\pi \sqrt {\alpha  - 1} }}{{4\sqrt 2 }}\frac{{\Delta _{tun} \tilde \Delta _{tun} }}{{\hbar eFv}}} \right),
\end{equation}
\begin{equation}
p_0  = \sqrt {\frac{{2\sqrt 2 }}{\pi }\frac{\alpha }{{\sqrt {\alpha  - 1} }}\frac{{\Delta _{tun} }}{{\tilde \Delta _{tun} }}\frac{{\hbar eF}}{v}}, 
\end{equation}
and
$\alpha  = \sqrt {1 + \left( {\frac{{2\gamma _1 }}{{\Delta _{tun} }}} \right)^2 }$. To avoid special functions, we extend the integration region up to infinite $p_{\perp}$ and obtain
\begin{equation}
J_{tun}  \approx \frac{{egkT}}{{2\pi ^{3/2} \hbar ^2 }}{\cal D}_0 p_0 \ln \left[ {\frac{{f_S \left( {E_v } \right)f_D \left( {E_c } \right)}}{{f_S \left( {E_c } \right)f_D \left( {E_v } \right)}}} \right].
\end{equation}

\begin{figure}[t]
\centering
\includegraphics[width=0.6\linewidth]{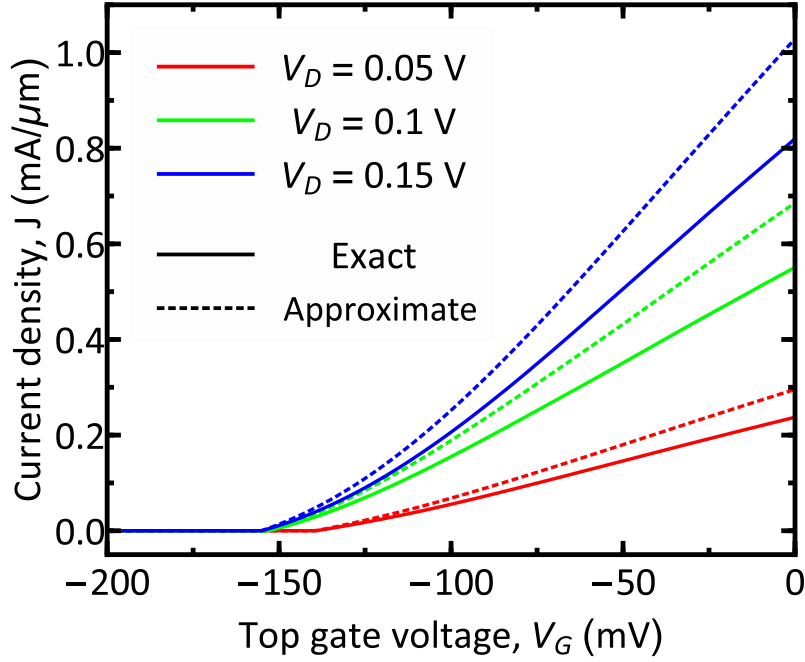}
\caption{Comparison of gate transfer characteristics at different drain bias obtained via numerical calculations (solid lines) and via approximate analytic equations (dashed lines). In the proposed GBL TFET tunneling occurs through varying bandgap, so we chose $\Delta_{tun}=\Delta_S$ in order to partially compensate the increase in current introduced by integrating up to infinite transverse momenta\label{ApproxCurrent} }
\end{figure}

Now we move on to the calculation of the thermionic current. Near the band edges the dispersion law is approximately parabolic:
\begin{equation}
p_{\max } \left( E \right) \approx p_{\min }  + \frac{{\Delta \gamma _1 }}{{2v^2 p_{\min } }}\sqrt { \pm \frac{E}{{\tilde \Delta }} - \frac{1}{2}},
\end{equation}
where plus sign corresponds to the conduction band and minus sign corresponds to the valence band. Taking into account that in Eq.~(\ref{thermionic}) the Fermi-Dirac distribution almost coincides with the Boltzmann distibution, we obtain
\begin{gather}
J_e  \approx \frac{{egkT}}{{2\pi ^2 \hbar ^2 }}\left( {p_{\min }^S  + \frac{{\Delta _S \gamma _1 }}{{4v^2 p_{\min }^S }}\sqrt {\frac{{\pi kT}}{{\tilde \Delta _S }}} } \right)\left[ {f_S \left( {E_c^S } \right) - f_D \left( {E_c^S } \right)} \right], \\
J_h  \approx \frac{{egkT}}{{2\pi ^2 \hbar ^2 }}\left( {p_{\min }^D  + \frac{{\Delta _D \gamma _1 }}{{4v^2 p_{\min }^D }}\sqrt {\frac{{\pi kT}}{{\tilde \Delta _D }}} } \right)\left[ {f_S \left( {E_v^D } \right) - f_D \left( {E_v^D } \right)} \right],
\end{gather}
where index "S" corresponds to source, and "D" corresponds to drain.

\section{Comparison with TFETs based on 2d semiconductors with parabolic bands}
In this section, we compare the dependencies of interband tunnel current on the band overlap in GBL and an equivalent 2D semiconductor without the 'Mexican hat' band structure. The equivalence of the materials implies equal values of band gap and barrier transparency in the electric field, while the band structure outside the gap is different. For the sake of analytical traceability, we assume the barrier transparency ${{\mathcal{D}}_{0}}$ to be weakly dependent on electron energy and gate voltage. The low-energy band structure of GBL is given by
Eqs.~(\ref{spectrum}). The electron dispersion of a parabolic-band material is
	\begin{equation}
	{{E}_{c}}\left( p \right)={E_G}/2+\frac{{{p}^{2}}}{2{{m}_{c}}}
	\end{equation}
in the conduction band,
	\begin{equation}
	{{E}_{v}}\left( p \right)=-{{E}_{G}}/2-\frac{{{p}^{2}}}{2{{m}_{v}}},
	\end{equation}
in the valence band. Here ${m_c}$ and ${m_{v}}$ are the effective masses for in the respective bands. The interband tunneling occurs mainly between conduction and light hole bands as the barrier transparency is small for heavy carriers. 
\begin{figure}[t]
\centering
\includegraphics[width=0.6\linewidth]{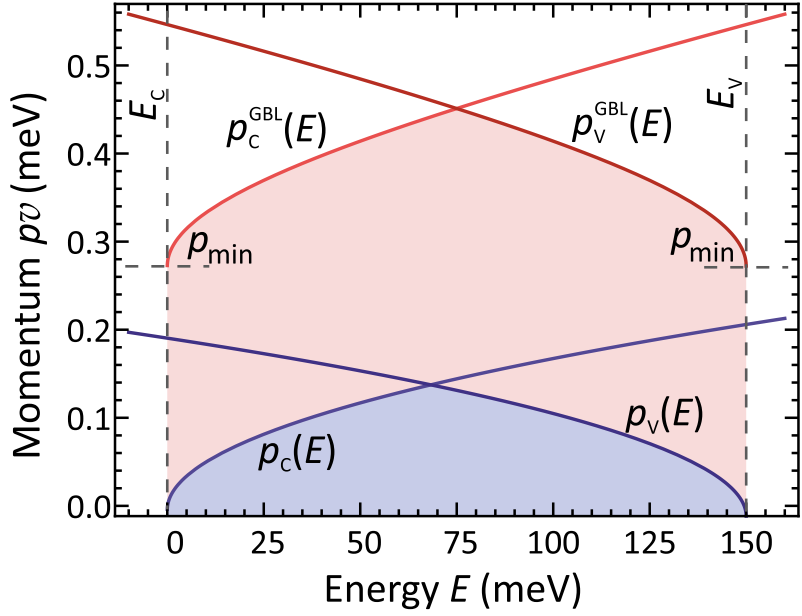}
\caption{\label{G7}Energy dependencies of electron momentum in the overlapping conduction and valence bands in a voltage-biased tunnel junction. Red curves are for graphene bilayer ($\tilde\Delta = 0.3$ eV), blue curves are for an equivalent parabolic-band 2D semiconductor. The filled area under the curves is proportional to the interband tunnel current.}
\end{figure}

Similarly to the derivation of Eq.~(2) of the main text, one can write down the current density in a parabolic-band 2d semiconductor
	\begin{equation}
	\label{Parband_Integral}
	J_{\rm{t, par}}\approx e\frac{g_s}{h^2}{{\mathcal D}_0}\int\limits_{{E_c}}^{{E_v}}{dE\left[ f_v(E)-{f_c}( E) \right]\int\limits_{0}^{{{p}_{\max }}\left( E \right)}{2dp}\approx }2e\frac{g_S}{h^2}{{\mathcal{D}}_0}\int\limits_{E_c}^{E_v}{p_{\max }(E) dE},
	\end{equation}
where $p_{\max}( E )$ is the maximum transverse momentum of electron with energy $E$ in the conduction and valence bands. The value of ${{p}_{\max }}(E)$ is the minimum of electron momenta at given energy in the conduction and valence bands:
	\begin{equation}
	{{p}_{\max }}\left( E \right)=\min \left\{ {p_c}(E),p_v(E) \right\}=\min \left\{ \sqrt{2m_c\left( E - E_c \right)},\sqrt{2{m_v}\left( E_v -E \right)} \right\}.
	\end{equation}
	
Graphically, the current $J_{\rm{t, par}}$ is proportional to the area under the inverse dispersion laws in the conduction and valence bands, this region is filled with blue in Fig.~\ref{G7}. The integral in (\ref{Parband_Integral}) is evaluated analytically with the result
	\begin{equation}
	\label{Parband_result}
	J_{\rm{t, par}}\approx e\frac{4\sqrt{2}}{3}\frac{{{g}_{S}}}{{{h}^{2}}}\sqrt{\frac{m_c m_v}{m_c+m_v}}{{\mathcal{D}}_0}{{\left( E_v-E_c \right)}^{3/2}}.
	\end{equation}

In graphene bilayer, the tunneling current density is given by
	\begin{equation}
	\label{GBL_Integral}
	J_{rm t}\approx 2e\frac{g_s g_v}{h^2}\left( 2{{\mathcal{D}}_{0}} \right)\int\limits_{E_c}^{E_v}{{{p}_{\max }}(E)dE},
	\end{equation}
where an extra factor of $g_v=2$ accounts for the valley degeneracy, and the factor of two in front of the barrier transparency is due to the two points in electron dispersion $E_c (p)$ with zero group velocity where an incident electron attempts to tunnel from the valence band. Again, ${{p}_{\max }}\left( E \right)=\min \left\{ {{p}_{c}}\left( E \right),{{p}_{v}}\left( E \right) \right\}$, where ${{p}_{c,v}}\left( E \right)$ are the dependencies of electron momentum on energy in the conduction and valence bands. Inverting the electron dispersion, one finds
	\begin{gather}
	p_c( E )=\frac{1}{v}\sqrt{(E-E_c)^2+\frac{\Delta^2}{4}+\sqrt{{(E-E_c)^{2}}( \gamma_1^2+{\Delta^2} )-\frac{\gamma _{1}^{2}{{\Delta }^2}}{4}}},\\
	p_v( E )=\frac{1}{v}\sqrt{(E_v-E)^2+\frac{\Delta^2}{4}+\sqrt{{(E_v-E)^2}( \gamma_1^2+{\Delta^2})-\frac{\gamma_1^2{\Delta^2}}{4}}}.
	\end{gather}

The energy integral (\ref{GBL_Integral}) for GBL cannot be evaluated analytically. However, the area under the dependence ${p_{\max }}\left( E \right)$ is naturally larger for graphene bilayer than for a typical narrow-gap semiconductor, not to mention an extra factor of four due to valley degeneracy and twofold tunneling. At small energies close to $E_c$, $p_c\left( E \right)\approx {{p}_{\min }}>\sqrt{2{{m}_{c}}\left( E-{{E}_{c}} \right)}$, while at large energies ${{p}_{c}}\left( E \right)\approx E/{v}$, which again lies above $\sqrt{2{{m}_{c}}\left( E-{{E}_{c}} \right)}$. In Fig. 5 of the main text, we numerically compare the current densities for GBL and 2d semiconductor with parabolic bands using Eqs.~(\ref{GBL_Integral}) and (\ref{Parband_result}). At the operating voltage of 150 mV, the current in GBL is 15 times larger than the current in a parabolic band semiconductor, of which the factor of 2 is due to the extra valley degeneracy, the factor of 2 is due to the two-fold tunneling, and the factor of 3.5 is due to the 'Mexican hat' dispersion.

\section{Estimate of the band tails in graphene bilayer}
The abruptness of tunnel switching is limited by the tunneling from the localized states in the band tails~\cite{Kane_Tails}. This mechanism can be alternatively interpreted as tunneling from the band tails, produced by the random dopant-induced potential fluctuations. In this section, we evaluate the density of states in graphene bilayer due to these fluctuations. In the quasi-classical limit, the 'true' density of states $\rho(E)$ is evaluated by integrating the bare density of states shifted by $eV$, $\rho_0(E-eV)$, timed by the probability of voltage fluctuation of magnitude $eV$
\begin{equation}
\label{TrueDoS}
\rho(E) = \frac{1}{\sqrt{2\pi \langle V^2\rangle}} \int\limits_{-\infty}^{+\infty}{\rho_0(E - e V) \exp\left(-\frac{V^2}{2\langle V^2\rangle} \right) dV }.
\end{equation}
In the above equation, $\left\langle {V^2} \right\rangle$ is the root-mean square voltage fluctuation,
\begin{equation}
\label{RMS_voltage}
\left\langle {V^2} \right\rangle =\frac{1}{S}\int{d{\bf x}{V^2}( \bf{x} )}=\frac{n_i}{{{\left( 2\pi  \right)}^2}}\int{d{\bf{q}}{{\left| {V_0}\left( {\bf{q}} \right) \right|}^2}},
\end{equation}
where $V_0({\bf q})$ is the Fourier transform of the potential produced by a single impurity and $n_i$ is the impurity density. To obtain a non-divergent result for fluctuations in two dimension, a finite distance from the charged impurities to the 2d plane has to be considered~\cite{Adam_SelfConsTransport}, in this case the Fourier transform of potential is
\begin{equation}
V_0({\bf q}) =  \frac{2 \pi e}{\kappa q \varepsilon(q) } e^{- q d_i}.
\end{equation}
Above, $\kappa$ is the background dielectric constant and $\varepsilon(q)$ is the dielectric function of graphene bilayer itself. For analytical traceability, the latter is taken in the Thomas-Fermi screening approximation~\cite{DasSarma_Screening}:
\begin{equation}
\varepsilon \left( q \right)=1+{q_s}/q,
\end{equation}
where the screening wave vector is calculated as
\begin{equation}
    q_s = \frac{2\pi e^2}{\kappa}\frac{1}{kT}\int_{-\infty}^{+\infty}{\rho (E) f(E) [1-f(E)]dE}.
\end{equation}
The integral (\ref{RMS_voltage}) is evaluated analytically with the result
\begin{equation}
\label{Vrms}
\left\langle {V^{2}} \right\rangle = 2\pi n_i \left(\frac{e}{ \kappa }\right)^2 K ( q_s d_i ),
\end{equation}
where $K\left( x \right)=-{e^{2x}}(2x+1)\text{Ei}(-2x)-1$. Equations (\ref{TrueDoS}) and (\ref{Vrms}) were used to calculate the density of states in the Fig.~5 of the main text numerically. The evaluation of DoS is, actually, an iterative procedure as the screening wave vector depends on the DoS itself. However, it converges quickly and the first approximation already is close to the final result.

\section{Modeling of the gate leakage}

\begin{figure}[t]
\centering
\includegraphics[width=0.6\linewidth]{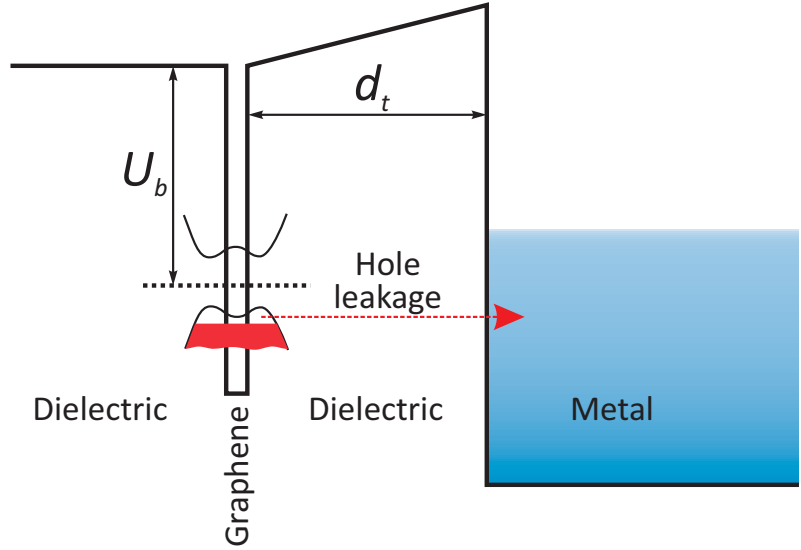}
\caption{A model band diagram of graphene bilayer -- dielectric -- metal system used for the evaluation of gate leakage current. Graphene bilayer is modelled as a delta-well with potential $U(z) = - \alpha \delta (z+d_t)$, where $\alpha$ is chosen to provide a correct work function from graphene to the insulating material. \label{BandsLeakage}}
\end{figure}

To estimate the leakage current from graphene bilayer to the metal gate, one has to multiply the charge density in graphene by the area of the gate and by the characteristic 'tunneling frequency' $\nu_t$. The latter can be derived from a lifetime of electron state bound to graphene bilayer in the presence of continuum of states representing the metal gate (Fig.~\ref{BandsLeakage}). 

The highest voltages are applied to the 'doping' gates, and the largest leakage is expected therein. An example of the band diagram corresponding to source doping gate is shown in Fig.~\ref{BandsLeakage}. Graphene bilayer is modelled as a delta-well $U(z) = -\alpha \delta(z+d_t)$. The parameter $\alpha = 2\hbar \sqrt{U_b/2m^*}$ is chosen to provide a correct value of the work function from graphene bilayer to the surrounding insulator material, $m^*$ is the electron effective mass in the insulator.

A simple wave function matching procedure for the potential profile of Fig.~\ref{BandsLeakage} leads to the dispersion equation for the quasi-bound states in graphene bilayer
\begin{equation}
\label{Leakage_disp}
    \left( \kappa -{{\kappa }_{B}} \right)\left( \kappa -ik_F \right)={{e}^{-2\kappa d}}{{\kappa }_{B}}\left( \kappa +ik_F \right),
\end{equation}
where ${\kappa }_{B} = \sqrt{2 m U_b}/\hbar$ is the decay constant of the wave function, $k_F$ is the electron wave vector in metal (approximately equal to the Fermi wave vector), and $\kappa = \sqrt{-2 m E}/\hbar$ is related to the sought-for state energy $E<0$. The solutions of Eq.~(\ref{Leakage_disp}) are complex, $E \approx -U_b + i \hbar \nu_t / 2$, where 
\begin{equation}
\label{Tun_freq}
    \nu_t = \frac{8 U_b}{\hbar} \frac{\kappa_B k_F}{\kappa_B^2+ k_F^2} \exp \left[ -2 \kappa_B d \right].
\end{equation}
For non-rectangular barriers, the exponent $e^{ -2 \kappa_B d }$ should be replaced with the quasi-classical transparency of the barrier separating graphene bilayer and the gate. The full leakage current density (per unit width of the channel) is evaluated as 
\begin{equation}
    J_g = e L_g \nu_ t (n_e + n_h),
\end{equation}
where $n_h$ is the density of holes induced at the source doping gate ($4.5 \times 10^{12}$ cm$^{-2}$ for $U_S = -0.6$ V) and $n_e$ is the electron density at the drain doping gate ($8 \times 10^{12}$ cm$^{-2}$ for $U_D =0.25$ V). To provide minimum tunneling frequency, we choose zirconium oxide as a gate dielectric due to its relatively large effective (tunneling) mass $m^* = 0.3 m_0$ and small electron affinity $\chi_{\rm{ZrO_2}} = 1.6$ eV leading to a quite large value of $U_b = \chi_{\rm{Gr}} - \chi_{\rm{ZrO_2}} = 2.9$ eV.

\end{document}